\documentclass{aa}

\usepackage{graphicx}
\usepackage[varg]{txfonts}
\usepackage{natbib}
\bibpunct{(}{)}{;}{a}{}{,}

\begin{document}

   \title{Decoupling of a giant planet from its disk in an inclined binary
   system}

   \author{G. Picogna\inst{1} \and F. Marzari\inst{2}}

   \institute{Institut f\"{u}r Astronomie und Astrophysik,
	      Universit\"{a}t T\"{u}bingen, Auf der Morgenstelle 10,
              72076 T\"{u}bingen, Germany\\
              \email{giovanni.picogna@uni-tuebingen.de}
         \and
              Dipartimento di Fisica, University of Padova,
              Via Marzolo 8, 35131 Padova, Italy
             }

   \date{Received 23 March 2015 / Accepted 20 August 2015}

   \titlerunning{Giant planets in inclined binaries}
   \authorrunning{G. Picogna and F. Marzari}

  \abstract
  {We explore the dynamical evolution of a planet embedded in a disk surrounding
   a star part of a binary system where the orbital plane of the binary is
   significantly tilted respect to the initial disk plane.
  }
  {Our aim is to test whether the planet remains within the disk and continues
   to migrate towards the star in a Type I/II mode in spite of the secular
   perturbations of the companion star.
   This would explain observed exoplanets with significant inclination respect
   to the equatorial plane of their host star.
  }
  {We have used two different SPH codes, \textsc{vine} and \textsc{phantom}, to
   model the evolution of a system star+disk+planet and companion star with
   time.
  }
  {After an initial coupled evolution, the inclination of the disk and that of
   the planet begin to differ significantly.
   The period of oscillation of the disk inclination, respect to the initial
   plane, is shorter than that of the planet which evolves independently after
   about $10^4\,\mbox{yr}$ following a perturbed N--body behavior.
   However, the planet keeps migrating towards the star because during its
   orbital motion it crosses the disk plane and the friction with the gas causes
   angular momentum loss.
  }
  {Disk and planet in a significantly inclined binary system are not dynamically
   coupled for small binary separations but evolve almost independently.
   The planet abandons the disk and, due to the onset of a significant mutual
   inclination, it interacts with the gas only when its orbit intersects the
   disk plane.
   The drift of the planet towards the star is not due to type I/II with the
   planet embedded in the disk but to the friction with the gas during the disk
   crossing.
  }

   \keywords{Protoplanetary disks --- Methods: numerical ---
    Planets and satellites: formation}
   \maketitle

\section{Introduction}
\label{intro}

  According to observational surveys, about half of solar--type stars reside in
  multiple stellar systems \citep{Raghavan2010} with the frequency declining to
  roughly about $30\%$ for less massive M stars \citep{Lada2006}.
  This frequency is suspected to be higher among young stars
  \citep{Reipurth2000} and the subsequent decay to the present fraction may be
  due to dynamical instability or gravitational encounters with other stars.
  The presence of circumstellar disks around both components of binaries does
  not seem to be strongly affected by the gravitational perturbations of the
  companion.
  Spatially resolved observations of disks in binaries in the Orion nebula
  cluster \citep{Daemgen2012} suggest that the fraction of circumstellar disks
  around individual components of binary systems is about $40\%$, only slightly
  lower than that for single stars (roughly $50\%$).
  On one side the secondary star perturbations may lead to disk truncation
  \citep{Artymowicz1994, Artymowicz1996, Bate2000} reducing its lifespan, on the other
  side the presence of a circumbinary disk may fed the truncated circumstellar disks
  through gas streams so that their dissipation timescale does not vary significantly
  compared to single stars \citep{monin07} and they have a chance to form planets.

  Up to date, $97$ exoplanetary systems have been discovered around multiple stars
  \citep{OEC}\footnote{\url{github.com/OpenExoplanetCatalogue/open_exoplanet_catalogue/}},
  triggering statistical comparisons with planets around single stars
  \citep{Desidera2007, Roell2012}.
  The analysis indicates that planetary masses increase for smaller stellar separation
  while for wide binaries the physical and orbital parameters appear very similar.
  In effect, planet formation in close binary configurations is a complex problem
  \citep[see][for a review]{Thebault2014} while it is expected that for large separations
  the effects of the binary companion on the planet growth may be less relevant.
  However, if the orbit of the companion star is significantly inclined respect to the
  initial protostellar disk plane, which is possibly co-planar with the star equator, this
  may lead to important dynamical consequences on the final orbit of the planet, in
  particular its inclination respect to the star equator.
  It has in fact been suggested \citep{Batygin2012} that the evolution of the
  protoplanetary disk under the perturbations of the binary may be responsible for the
  observed spin--orbit misalignment of some exoplanets.
  According to \citet{Triaud2010} and \citet{Albrecht2012} about $40\%$ of hot Jupiters
  have orbits significantly tilted respect to the equatorial plane of the star.
  A fundamental requirement of the model of \citet{Batygin2012} is that the planet resides
  within the disk during its evolution.
  In this way it would continue its migration by tidal interaction with the disk and, at
  the same time, it will follow the inclination evolution of the disk.
  Once the disk is dissipated its migration will stop, while the evolution of the
  inclination will continue, though in a different fashion, until the binary companion is
  possibly stripped away leaving the planet on an inclined orbit.

  Previous studies seem to suggest that indeed a giant planet is forced to evolve within
  its birth disk even in presence of external perturbing forces.
  Any relative inclination between the planet and the disk plane, induced either by a
  planet--planet scattering event or by resonances \citep{thomlis03}, is quickly damped by
  the disk leading to a realignment of the planet orbit with the disk plane
  \citep{Cresswell2007, Marzari2009, Bitsch2011, Rein2012, Teyssandier2013}.
  However, the precession rate of the planetary orbit due to the interaction with the
  binary star is faster than that of the disk.
  As a consequence, if the damping of the disk on the planet orbit is not strong enough to
  keep it within the disk plane, the planet will evolve independently under the secular
  perturbations of the secondary star and develop a significant relative inclination
  respect to the disk.
  The assumed type II migration will not occur in this case since the planet is not
  embedded in the disk and cannot open a gap.

  In a recent paper, \citet{Xiang2014} have investigated the evolution of a disk and
  massive planet under the influence of an inclined binary companion.
  According to their SPH simulations, the planet and disk maintain approximate coplanarity
  during the evolution of the system and the planet would quickly migrate towards the star
  and stop on a close orbit.
  This findings seem to confirm the model described in \citet{Batygin2012}  for
  small separations between the binary components.
  To further test these findings, we have performed numerical simulations with a setup
  similar to that adopted in \citet{Xiang2014} and used two different SPH codes:
  \textsc{vine} \citep{Wetzstein2009,Nelson2009} and \textsc{phantom}
  \citep{Price2010,Lodato2010}.
  Our runs extend over a much longer timescale and we adopted a higher resolution.
  The simulations show that the initial coplanarity is maintained only for a limited
  amount of time and that the planet definitively detaches from the disk plane as soon as
  the disk completes a quarter of its precession period.
  The secular perturbations of the binary companion in an inclined orbit overcome the
  damping force of the disk and the planet evolves independently of the disk.
  A significant mutual inclination develops between the disk and the planet orbital plane
  but this does not halt the orbital migration.
  When the planet crosses the disk it is still affected by the gas via dynamical friction
  and its inward  drift continues even if at a slower rate compared to that computed by
  \citet{Xiang2014} in the initial stages of the system evolution.

  In Sec.~\ref{sec:theo} we study the theoretical model of the precession rates in a
  star+disk+planet--star system and compare evolution timescales.
  In Sec.~\ref{sec:model} we describe in detail the model setup of the numerical
  simulations.
  Then, in Secs.~\ref{sec:init45} and \ref{sec:init60} we show the results of the long
  term high resolution models for different binary inclinations, and in
  Sec.~\ref{sec:concl} we discuss our results and their implications.

\section{Relevant timescales}
\label{sec:theo}

  We compare in this section the timescales of precession of the angular momentum of the
  planet computed within a pure N--body problem and that estimated for the disk.
  This may give important hints on the forces that may try to separate the planet from the
  disk and give a better insight on the results of our SPH simulations.
  In computing the precession timescales, we consider a planet of mass $M_\mathrm{p}$
  initially on a circular orbit around a star of mass $M_\star$ at a semimajor axis $a$,
  with a distant binary companion of mass $M_\mathrm{b}$, semimajor axis $a_\mathrm{b}$,
  inclination $i_\mathrm{B}$, and eccentricity $e_\mathrm{b}$ which we assume to be $0$ at
  the beginning.
  The perturbations of the companion star are strong in particular for inclinations higher
  than $i_\mathrm{B} \sim 39^o$ when the Kozai--Lidov mechanism \citep{kozai62, lidov62}
  forces wide oscillations of both the inclination and eccentricity of the planet whose
  values are tied by the conservation of
  $L_\mathrm{z} = \sqrt{(1-e^2)}\cos(i_\mathrm{p})$.
  The maximum eccentricity reachable by the planet is
  $e_\mathrm{max}\simeq\sqrt{1-5/3\cos^2{\theta_\mathrm{lb}^0}}$.
  However, short range forces like General Relativity and tidal distortions
  (which we neglect here), and the dumping effect of the protoplanetary disk, tend to
  reduce this value.

  The Kozai cycles occur at a characteristic period given by \citep{kiseleva98}
  \begin{align}
    \label{eq:PKoz}
    P_\mathrm{KZ} &\simeq \frac{P_\mathrm{b}^2}{P_\mathrm{p}}(1-e_\mathrm{b}^2)^{3/2}
    \frac{M_\star+M_\mathrm{b}}{M_\mathrm{b}} \nonumber \\
    &=\frac{2\pi}{\Omega_\mathrm{p}}
    \left(\frac{a_\mathrm{b}}{a}\right)^3\left(\frac{M_\star}{M_\mathrm{b}}\right)
    (1-e_\mathrm{b}^2)^{3/2}
  \end{align}
  \citep[see also][]{ford2000} where $P_\mathrm{p}$ and $P_\mathrm{b}$ are the orbital
  periods of the planet and binary.
  The dynamical evolution of the planet can also be viewed from the point of view of the
  orbital angular momentum.
  According to \citet{Storch2014}, the planetary orbital angular momentum vector precesses
  around the binary axis ($\vec{\hat{L}}_\mathrm{b}$) at a rate that, in the absence of
  tidal dissipation, is approximately given by
  \begin{equation}
    \label{eq:PrecKoz}
    \Omega_\mathrm{pb}\simeq\frac{3\pi}{2}P_\mathrm{KZ}^{-1}
    \cos{\theta_\mathrm{pb}^0}\sqrt{1-e_0^2}\left[1-2\left(\frac{1-e_0^2}{1-e^2}\right)
    \frac{\sin^2{\theta_\mathrm{pb}^0}}{\sin^2{\theta_\mathrm{pb}}}\right]
  \end{equation}
  where $e_0$ is the initial eccentricity of the planet, $\theta_\mathrm{pb}$ is the
  inclination of the planetary orbit respect to the binary orbit, and
  $\theta_\mathrm{pb}^0$ is its initial value.
  The precession of the orbital angular momentum approximately translates into an
  oscillation of the planet inclination respect to the initial plane which is the disk
  plane.
  The timescales estimated by eq.~\ref{eq:PKoz} and \ref{eq:PrecKoz} are then
  comparable.

  Also the disk is affected by the binary companion gravity and the disk axis
  $\vec{\hat{L}}_\mathrm{d}$ precesses around the binary axis $\vec{\hat{L}}_\mathrm{b}$,
  assuming that it satisfies the condition for a coherent behavior as a solid body, with
  a period given by
  \begin{align}
    P_\mathrm{d}^\mathrm{prec} &\simeq P_\mathrm{out}\left(\frac{a_\mathrm{b}}
    {r_\mathrm{out}}\right)^3
    \left(\frac{M_\star}{M_\mathrm{b}}\right)\frac{1}{K\cos{\theta_\mathrm{db}}}
    \nonumber \\
    &\simeq 2\pi\left(\frac{a_\mathrm{b}^6}{GM_\star r_\mathrm{out}^3}\right)^{1/2}
    \left(\frac{M_\star}{M_\mathrm{b}}\right)\frac{1}{3/8\cos{\theta_\mathrm{db}}}
  \end{align}
  \citep[see][]{Batebonn2000,Lai2014}, where $r_\mathrm{out}$ and $P_\mathrm{out}$ are
  the radius and Keplerian period of the outer disk edge respectively, and
  $\theta_\mathrm{db}$ is the inclination of the disk respect to the orbital plane of
  the binary.
  The relative precession rate is given by
  \begin{equation}
    \Omega_\mathrm{db} \simeq -\frac{3}{8}\left(\frac{GM_\star r_\mathrm{out}^3}
    {a_\mathrm{b}^6}\right)^{1/2}\left(\frac{M_\mathrm{b}}{M_\star}\right)
    \cos{\theta_\mathrm{db}}
  \end{equation}
  Even in this case, the precession can be translated in a periodic oscillation of the
  disk inclination respect to the initial plane which can be assumed to have been the
  equatorial plane of the primary star.

  If we now compare the two precession rates, that of the planet and that of the disk,
  we find that the initial ratio between them, under the assumption that the planet formed
  in the mid--plane of the disk
  ($\theta_\mathrm{pb}=\theta_\mathrm{pb}^0=\theta_\mathrm{db}$) in a circular orbit
  ($e=e_0=0$), is given by
  \begin{align} \label{Pratio}
    \left(\frac{P_\mathrm{d}^\mathrm{prec}}{P_\mathrm{p}^\mathrm{prec}}
    \right)_\mathrm{ini} =
    \left(\frac{\Omega_\mathrm{pb}}{\Omega_\mathrm{db}}\right)_\mathrm{ini} &\simeq
    4\pi P_\mathrm{KZ}^{-1}\left(\frac{a_\mathrm{b}^6}{GM_\star
    r_\mathrm{out}^3}\right)^{1/2} \frac{M_\star}{M_\mathrm{b}} \nonumber \\
    &= 2\left(\frac{a}{r_\mathrm{out}}\right)^{3/2} = 2\left(\frac{a}
    {a_\mathrm{b}/3}\right)^{3/2}
  \end{align}
  where in the last step we assumed that the binary truncates by tidal interaction the
  disk maximum radius to $1/3$ of the binary separation, which is a good approximation
  for binary separations less than $\sim 300\,\mbox{au}$ \citep{Artymowicz1994}.

  For $t\neq t_0$ and $e_0 = 0$ the ratio is evolving as
  \begin{equation}
    \frac{P_\mathrm{d}^\mathrm{prec}}{P_\mathrm{p}^\mathrm{prec}}
    \simeq 2\frac{\cos{\theta_\mathrm{lb}^0}}
    {\cos{\theta_\mathrm{db}}}\left(\frac{a}{a_\mathrm{b}/3}\right)^{3/2}
    \left[2\left(\frac{1}{1-e^2}\right)\frac{\sin^2{\theta_\mathrm{lb}^0}}
    {\sin^2{\theta_\mathrm{lb}}}-1\right]
  \end{equation}

  If we consider a typical case of an equal mass binary made of solar--type stars on a
  circular orbit and separated by $100\,\mbox{au}$ and a planet orbiting around the
  primary with $a=5\,\mbox{au}$ and embedded in a disk with outer radius of
  $30\,\mbox{au}$, from eq.~\ref{Pratio} the two period differs by about a factor ten.
  This is a large difference and the disk damping must overcome the tendency of the planet
  to evolve on a slower timescale and then out of the disk plane.

\section{Model description}
\label{sec:model}
  In this section we briefly describe the two numerical algorithms we have used to model
  the system planet+disk+binary companion and outline their differences which might affect
  long term results.

  \subsection{SPH codes and model setup}
  \label{sec:code}

    To model the evolution of a Jupiter--size planet embedded in the disk surrounding the
    primary star of an inclined binary system we have used two different SPH codes:
    \begin{itemize}
      \item \textsc{vine} \citep{Wetzstein2009,Nelson2009}, which is a hybrid N--body/SPH
	code, updated to improve momentum and energy conservation as described in
        \citet{Picogna2013},
      \item \textsc{phantom} \citep{Price2010,Lodato2010}, a modern SPH code which models
	the massive bodies as sink particles \citep{Bate1995}.
    \end{itemize}
    Both these codes solve the hydrodynamic equations in the Lagrangian approach by
    replacing the fluid with a set of particles \citep[see][for a review]{Price2012}.

   \vskip 0.5 truecm
   \par\noindent\emph{Equation of state---}
    A locally isothermal equation of state, similar to that described in
    \citet{Peplinski2008}, is adopted in all simulations
    \begin{equation}
      c_\mathrm{s} = \frac{h_\mathrm{s}r_\mathrm{s}h_\mathrm{p}r_\mathrm{p}}
        {[(h_\mathrm{s}r_\mathrm{s})^n+(h_\mathrm{p}r_\mathrm{p})^n]^{1/n}}
        \sqrt{\Omega_\mathrm{s}^2+\Omega_\mathrm{p}^2},
    \end{equation}
    where $r_\mathrm{s}$, $r_\mathrm{p}$ are the distances of the fluid element from the
    primary star and the planet, respectively, and $h_\mathrm{s}$, $h_\mathrm{p}$ are the
    circumstellar and circumplanetary disk aspect ratios.
    $\Omega_\mathrm{s}$ and $\Omega_\mathrm{p}$ are the circular Keplerian orbits of a
    fluid element around the star and planet, and $n=3.5$ is a non--dimensional parameter
    chosen to smoothly join the equation of state near the planet ($c_\mathrm{s} =
    h_\mathrm{p}r_\mathrm{p}\Omega_\mathrm{p}$ for $r_\mathrm{p} << r_\mathrm{s}$) with
    that of a flat circumstellar disk ($c_\mathrm{s} = h_\mathrm{s} r_\mathrm{s}
    \Omega_\mathrm{s}$).

    For the simulations in this work we choose a disk aspect ratio of
    $h_\mathrm{s}=0.037$, and a circumplanetary scale height of $h_\mathrm{s}=0.6$.

   \vskip 0.5 truecm
   \par\noindent\emph{Accretion---}
    \textsc{vine} treats the stars and planet as N--bodies.
    The mass accretion has been modelled for the stars inside a radius of
    $0.5\,\mbox{au}$, but not for the planet.

    In the simulations performed with the \textsc{phantom} code, the planet (and stars)
    are all modelled as Lagrangian sink particles \citep{Bate1995}.
    A sink particle is evolved as an SPH particle, but it experiences only the
    gravitational force.
    A gas particle that comes within its accretion radius,
    $R_\mathrm{acc}=0.075\,\mbox{au}$, can be accreted if
    \begin{itemize}
      \item it is inside the Hill sphere of the sink particle,
      \item its specific angular momentum is less than that required to form a circular
        orbit at the accretion radius,
      \item it is more bound to the candidate sink particle than to any other sink
        particle.
    \end{itemize}

   \vskip 0.5 truecm
   \par\noindent\emph{Viscosity---}
    An artificial viscosity term is introduced in SPH codes in order to correctly model
    shock waves that inject entropy into the flow over distances that are much shorter
    than a smoothing length, and to simulate the evolution of viscous disks.
    The term broadens the shock over a small number of smoothing lengths and correctly
    resolves it ensuring at the same time that the Rankine--Hugoniot equations are
    satisfied.
    In this way it prevents discontinuities in entropy, pressure, density, and velocity
    fields.

    The implementation of the viscosity term is slightly different in the two numerical
    codes.
    \textsc{vine} adopts the standard formulation introduced by \citet{Monaghan1983}
    where, for approaching particles ($\vec{v}_{ab}\cdot \vec{\hat{r}}_{ab}\leq0$), an
    artificial viscosity term is introduced in the momentum and energy equations
    \begin{equation}
      \Pi_{ab}=
        \begin{cases}
          \frac{(-\alpha c_{ab}\mu_{ab} + \beta \mu_{ab}^2)}{\rho_{ab}}
          	  & \vec{v}_{ab}\cdot \vec{\hat{r}}_{ab}\leq 0 \\
          0   & \vec{v}_{ab}\cdot \vec{\hat{r}}_{ab}> 0
        \end{cases}
    \end{equation}
    where $\mu_{ab}$ is a velocity divergence term
    \begin{equation}
      \mu_{ab}=\frac{h \vec{v}_{ab}\cdot \vec{r}_{ab}}{\vec{r}_{ab}^2+\eta^2h^2}
    \end{equation}
    with $\eta^2=0.01$.
    This implementation conserves total linear and angular momentum and vanishes for
    rigid body rotation.
    $\alpha$ is the linear term (bulk viscosity) and it dissipates kinetic energy as
    particles approach each other to reduce subsonic velocity oscillations following a
    shock.
    $\beta$ is the quadratic term (von Neumann--Richtmyer like viscosity), which converts
    kinetic energy to thermal energy preventing particle mutual penetration in shocks.

    \textsc{phantom} uses a more general formulation of dissipative terms
    \citep{Monaghan1997}.
    It is built on an analogy with Riemann solvers, where the dissipative terms of
    conservative variables (density, specific momentum, and energy), that experience a
    jump across a shock front are multiplied by eigenvalues similar to signal velocities
    ($v_\mathrm{sig}$).
    The viscosity term for the momentum equation becomes
    \begin{equation}
      \Pi_{ab}=\frac{1}{2}\frac{\alpha v_\mathrm{sig}
    	\vec{v}_{ab}\cdot\vec{\hat{r}}_{ab}}{\rho_{ab}}
    \end{equation}
    where the signal velocity is given by
    \begin{equation}
      v_\mathrm{sig}=
      \begin{cases}
        c_{\mathrm{s},a}+c_{\mathrm{s},b} - \beta\vec{v}_{ab}\cdot
      	  \vec{\hat{r}}_{ab} & \vec{v}_{ab}\cdot \vec{\hat{r}}_{ab}\leq 0 \\
        0 & \vec{v}_{ab}\cdot \vec{\hat{r}}_{ab} > 0
      \end{cases}
    \end{equation}
    Regarding the energy evolution an additional signal velocity $v_\mathrm{sig,u}$ is
    adopted, and the viscosity term is defined as
    \begin{equation}
      \left(\frac{du}{dt}\right)_{\mathrm{diss},a} = -\sum_b \frac{m_{b}}{\rho_{ab}}
    	\left[\frac{1}{2}\alpha v_\mathrm{sig}(\vec{v}_{ab}\cdot\vec{\hat{r}}_{ab})^2+
    	\alpha_\mathrm{u}v_\mathrm{sig,u}u_{ab}\right]\vec{\hat{r}}_{ab}\cdot
    	\nabla_{a}W_{ab}
    \end{equation}
    where the signal velocity for the energy jumps is chosen to be
    $v_\mathrm{sig,u}=|\vec{v}_{ab}\cdot\vec{\hat{r}}_{ab}|$ as in \citet{Wadsley2008}.

    In order to properly compare these two different approaches and derive a Shakura \&
    Sunyaev--like viscosity value for the different simulations, we adopted the relations
    given by \citet{Meru2012}
    \begin{equation} \label{alp1}
      \alpha_\mathrm{SS} = \frac{1}{20}\alpha\frac{h}{H} + \frac{3}{35\pi}\beta
      \left(\frac{h}{H}\right)^2
    \end{equation}
    for the \citet{Monaghan1983} formalism, and
    \begin{equation} \label{alp2}
      \alpha_\mathrm{SS} = \frac{31}{525}\alpha\frac{h}{H} + \frac{9}{70\pi}\beta
      \left(\frac{h}{H}\right)^2
    \end{equation}
    for the \citet{Monaghan1997} one, where $h$ is the averaged smoothing length, and $H$
    is the disk scale height.

    The initial average value of $\alpha_\mathrm{SS}$ for our \textsc{vine} run is $0.02$,
    similar to that in \citet{Xiang2014}.
    About twice larger is the $\alpha_\mathrm{SS}$ in the \textsc{phantom} run with an
    average initial value of $0.04$.

   \vskip 0.5 truecm
   \par\noindent\emph{Initial conditions---}
    The scenario we have modelled includes a binary system composed of $2$ equal mass
    stars $m_\mathrm{p} = m_\mathrm{s} = 1\ M_\sun$.
    One of them (which we will call primary) harbors a protoplanetary disk with a Jupiter
    mass planet $M_\mathrm{p} = 1\ M_\mathrm{J}$ embedded in it on an orbit with initial
    semimajor axis $a_\mathrm{p} = 5\,\mbox{au}$.
    Different inclinations between the two components of the binary has been studied, in
    a reference frame centered onto the primary star.
    The disk is modelled with $600\,000$ SPH particles, about $3$ times the number used
    by \citet{Xiang2014}, and it extends from $0.5$ to $30\,\mbox{au}$, with a surface
    density profile
    \begin{equation}
      \label{eq:surfdens}
      \Sigma  = \Sigma_0 r^{-0.5}
    \end{equation}
    where $\Sigma_0$ is defined such as the total disk mass is $M_\mathrm{D} = 0.01\
    M_\sun$.

    The choice of the initial parameters are such that the disk will precess as a solid
    body.
    In fact, according to \citet{PapaTer1995} and \citet{Larwood1996}, the condition for
    this behavior is
    \begin{equation}
      \label{eq:period}
      P_\mathrm{out} / P_\mathrm{d}  \leq H/R
    \end{equation}
    In our initial setup this condition is fully satisfied since $H/R$ is $0.037$ while
    the ratio between $\omega_\mathrm{p}/\Omega_\mathrm{d}$ is approximately $0.01$.
    This is confirmed also by the calculations of \citet{Xiang2013}.
    As reference frame we adopt the initial plane of the disk and planet orbit while the
    binary orbit has different initial inclinations defined respect to this plane.
    The initial semimajor axis of the companion star is $a_\mathrm{b} = 100\,\mbox{au}$
    and it is set on a circular orbit.

\section{Initial mutual inclination of $45^{\circ}$}
\label{sec:init45}

  The model where the inclination between the binary orbit and the initial disk+planet
  plane is $45^{\circ}$ is assumed as standard model.
  The simulation with \textsc{vine} was halted after $11\,600\,\mbox{orbits}$ of the
  planet ($125\,000\,\mbox{yr}$) while that with \textsc{phantom} was stopped after about
  half of that period ($60\,000\,\mbox{yr}$).
  The run with \textsc{vine} required approximately $8$ months of CPU on a $24$ processor
  machine while that with \textsc{phantom} $4$ months on a $32$ processor machine.
  In both simulations we find that in the first $6\,000\,\mbox{yr}$ the behavior is
  similar to that described in \citet{Xiang2014} but, soon after, the dynamical evolution
  substantially changes.
  The inclination of the planet $i_\mathrm{p}$, that initially follows closely that of
  the disk $i_\mathrm{d}$, decisively departs from $i_\mathrm{d}$ and grows at a lower
  rate.
  In Fig.~\ref{INC45f} both $i_\mathrm{p}$ and $i_\mathrm{d}$ are shown as a function of
  time for the run with \textsc{vine} and \textsc{phantom}.
  While the periodic oscillations of $i_\mathrm{d}$ continue with an almost unaltered
  frequency of about $2 \times 10^4\,\mbox{yr}$, the planet inclination $i_\mathrm{p}$
  grows on a much longer timescale reaching a maximum after about
  $8 \times 10^4\,\mbox{yr}$.
  There are clear indications of interaction between the planet and the disk in
  particular when the two inclinations are comparable like after $2.5 \times 10^4$,
  $5 \times 10^4$ etc.
  At these times the planet crosses the disk and it interacts with its gas particles.

\begin{figure}[hpt]
    \resizebox{.9\hsize}{!}{\includegraphics{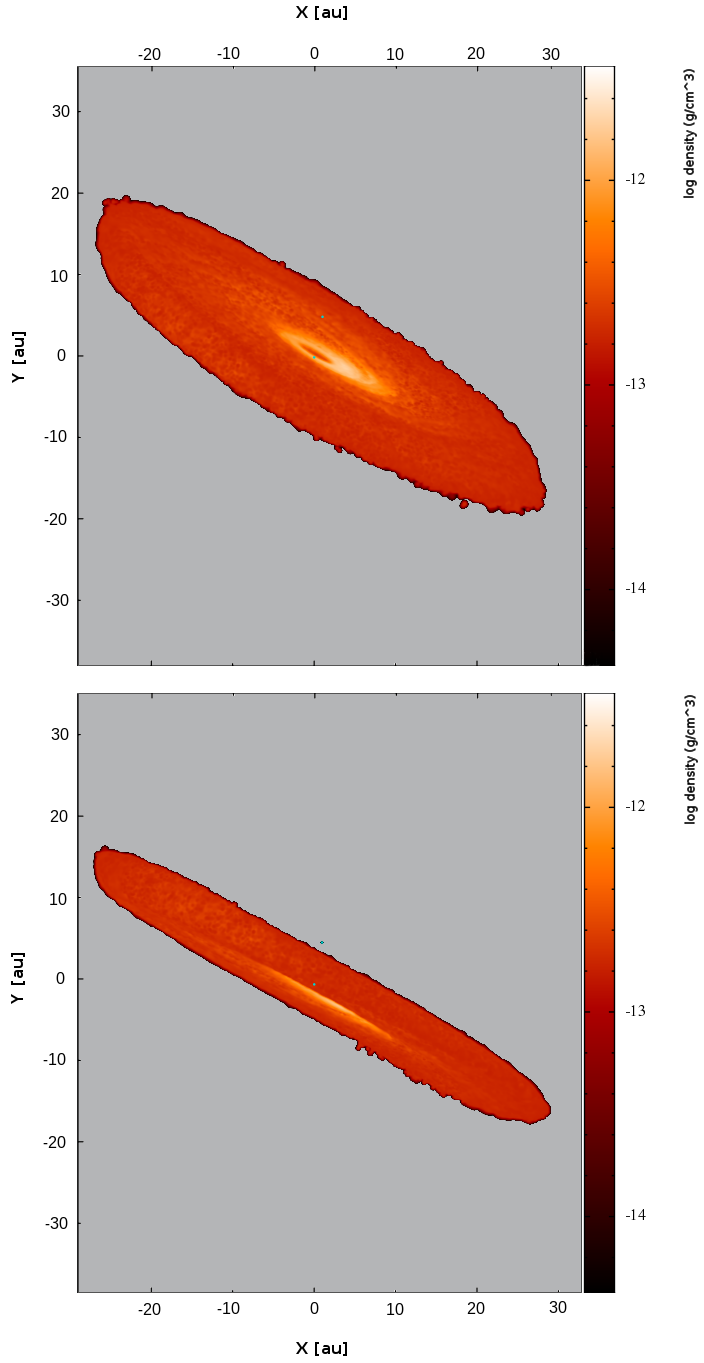}}
    \caption{\label{vinedisk}
    3D rendering of the disk + planet system in the \textsc{vine} run.
    The plot is created through a ray tracing process, from an observer point of view
    placed at $~\sim 150\,\mbox{au}$ from the disk.
    The disk is shown at an optical depth of $\sim 2$, defined as the cross section
    per SPH particle mass.}
  \end{figure}

\begin{figure}[hpt]
    \resizebox{\hsize}{!}{\includegraphics[height=8.5truecm]{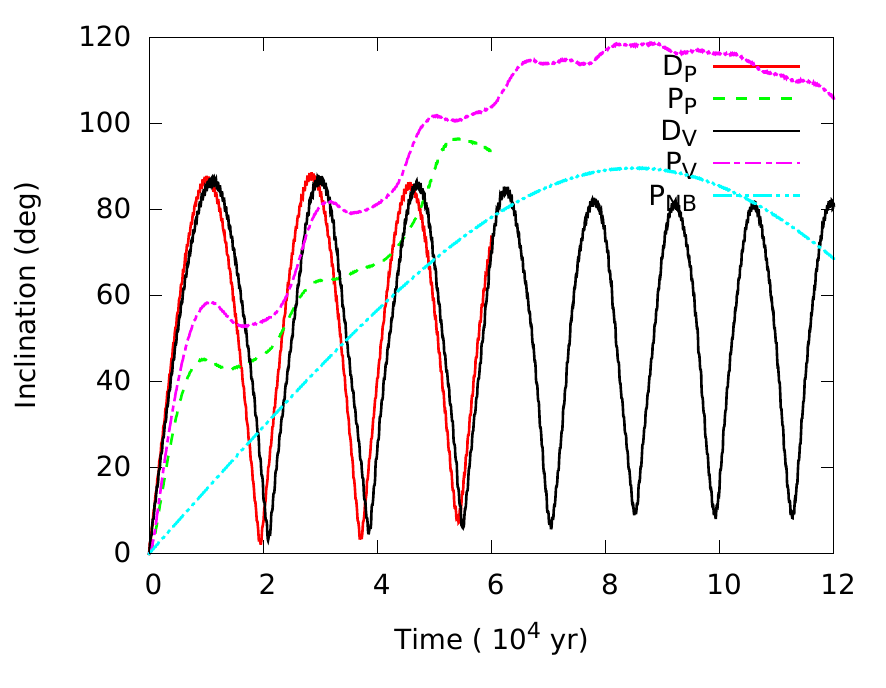}}
    \caption{\label{INC45f}
    Evolution of the disk and planet inclination respect to the initial disk+planet
    plane in the two runs with \textsc{vine} and \textsc{phantom}, respectively.
    The dashed magenta (\textsc{vine}) and green (\textsc{phantom}) lines show the planet
    inclination $i_\mathrm{p}$ while the continuous black (\textsc{vine}) and red
    (\textsc{phantom}) line illustrates the disk inclination $i_\mathrm{d}$.
    After about $6\,000\,\mbox{yr}$, the planet evolution departs  from that of the disk
    in both simulations and $i_\mathrm{p}$ grows at a slower rate compared to
    $i_\mathrm{d}$.
    The light blue dashed--dotted line illustrates the evolution of the planet inclination
    in a pure N--Body problem Star--Planet--Companion star.}
  \end{figure}

  There are some minor differences between the \textsc{vine} and the \textsc{phantom}
  models possibly related to the different handling of viscosity and accretion onto the
  planet.
  A circumplanetary disk develops in the initial phase of the run with \textsc{vine} when
  $i_\mathrm{p}\sim i_\mathrm{d}$ but later on when the planet departs from the disk and
  periodically crosses it with high mutual inclination it is dissipated.
  In both simulations dynamical behavior of the planet shows a clear competition between
  the gravitational force of the binary and the interaction with the disk, but the binary
  secular perturbations finally dominate.
  The inclination of the planet roughly follows a pure N--body trend, shown as a dotted
  light--blue line in Fig.~\ref{INC45f}, even if $i_\mathrm{p}$ is strongly perturbed by
  the disk.
  The detachment between the planet and the disk can be seen also in Fig.~\ref{vinedisk},
  where a 3D rendering of the system is displayed at an optical depth of $\sim$ 2.

  In Fig.~\ref{INC45f} both simulations show a slow damping of the
  inclination oscillations, a behavior already suggested by
  \cite{martin2014}.

\begin{figure}[hpt]
    \resizebox{\hsize}{!}{\includegraphics{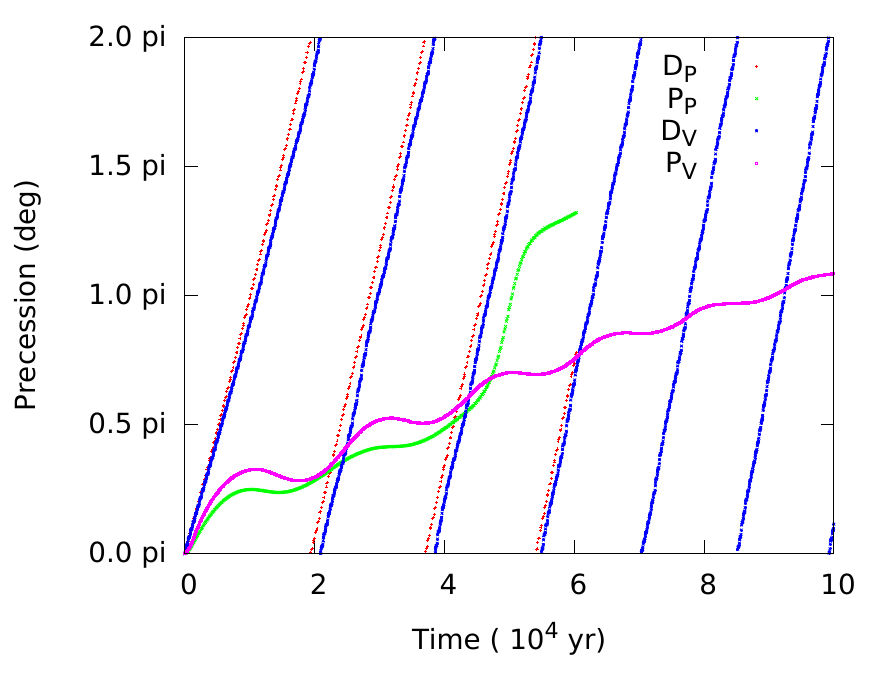}}
    \caption{\label{PREC}
    Evolution of the disk and planet precession angle respect to the initial disk+planet
    plane, as defined in eq.~\ref{eq:prec}.
    The dashed green line shows the planet precession angle $\beta_\mathrm{p}$ while the
    continuous red line illustrates the disk's one $\beta_\mathrm{d}$.}
  \end{figure}

  As discussed in Sec.~\ref{sec:theo}, the distinct evolution of the planet and disk is
  related to their different precession timescales around the binary axis which translates
  in the inclination evolution shown in Fig.~\ref{INC45f}.
  As an additional indication of the different evolution of the planet respect to the disk
  in Fig.~\ref{PREC} we illustrate the evolution of the precession angle of the disk
  $\mathbf{L}_\mathrm{D}$ and planet $\mathbf{L}_\mathrm{p}$ angular momentum around that
  of the binary star $\mathbf{L}_\mathrm{B}$, calculated respect to the initial orbital
  plane \citep{Xiang2014,Larwood1996}
  \begin{equation}\label{eq:prec}
    \cos{\beta_\mathrm{p}}=\frac{\mathbf{L}_\mathrm{D}\times\mathbf{L}_\mathrm{B}}
    {|\mathbf{L}_\mathrm{D} \times \mathbf{L}_\mathrm{B}|}\cdot\mathbf{u},
  \end{equation}
  where $\mathbf{u}$ is the fixed unit reference vector for the initial orbital plane.
  The evolution timescale of the planet precession rate is $\sim 10$ times slower than
  that of the disk, confirming the theoretical prediction.

\begin{figure}[hpt]
    \resizebox{\hsize}{!}{\includegraphics{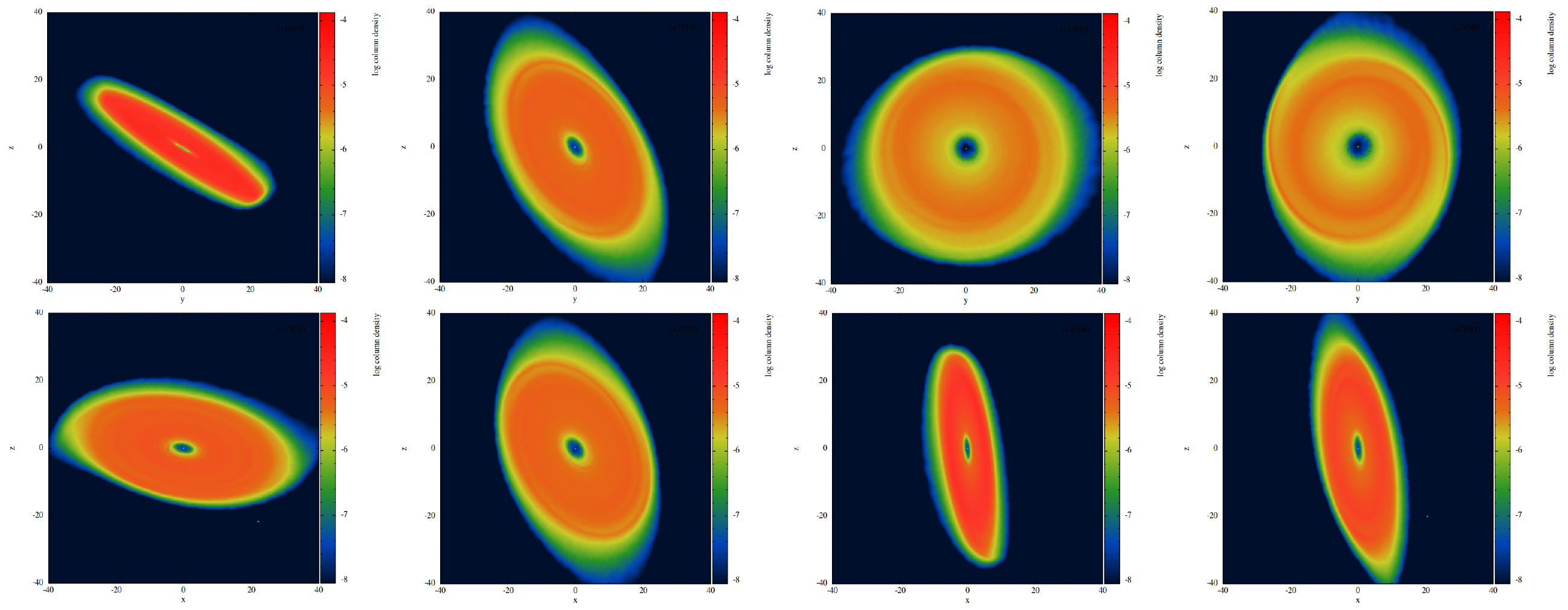}}
    \caption{\label{warp45}
    Disk shape shown every $300$ planetary orbits when the inclination of the
    planet is decoupling from that of the planet. The disk warping is negligible.}
  \end{figure}

\begin{figure}[hpt]
    \resizebox{\hsize}{!}{\includegraphics{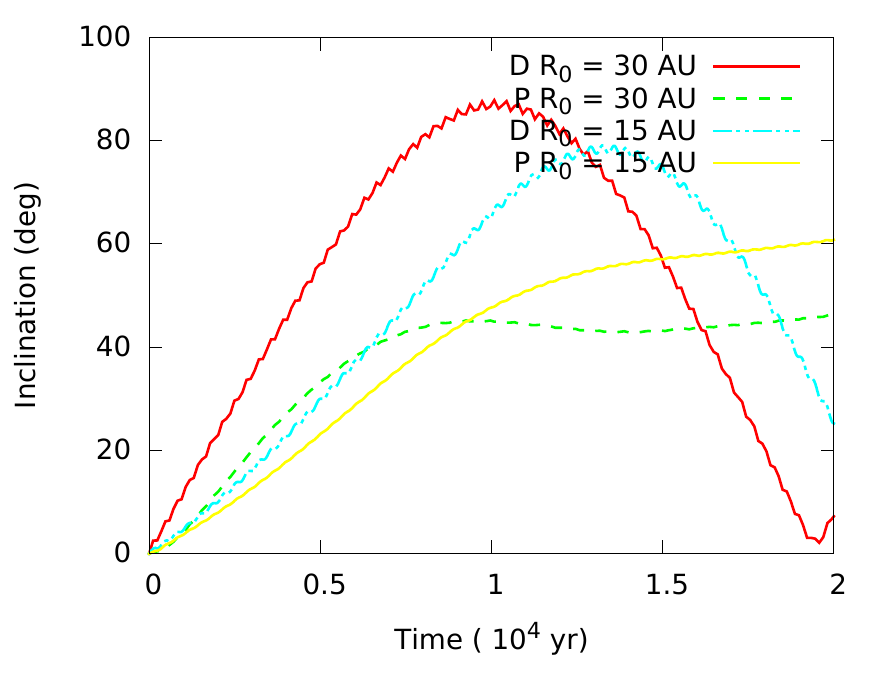}}
    \caption{\label{disk15}
    Comparison of the planet and disk inclination evolution in
    the standard case, where the disk extends to $30\,\mbox{au}$, and a
    model with a smaller disk, extending to $15\,\mbox{au}$.
    No warping is expected in the latter case but the planet
    decouples from the disk as in our standard case with
    a larger disk.}
  \end{figure}

  To test whether a potential warping of the disk due to binary perturbations is
  fully responsible for the decoupling of the planet from the disk in our
  models, as described in \citep{terquem2013}, we checked the disk shape every
  $300$ planetary orbits.
  In Fig.~\ref{warp45} we show the disk evolution at different times and there is no
  evidence of a marked warping.
  In addition, we also performed a simulation where a smaller disk was considered.
  In Fig.~\ref{disk15} the PHANTOM code was used to model the evolution of a disk with the
  same mass as our standard disk but extending only to $15\,\mbox{au}$ in radial distance.
  The warping for such a disk, being less affected by the binary perturbations extending
  well within the tidal truncation radius \citep{Artymowicz1994}, is expected to be
  negligible.
  Even in this case the planet decouple from the
  disk in a way very similar to that shown in  Fig.~\ref{INC45f} even if on
  slightly different timescales due to the different disk configuration
  and density. Initially, the smaller disk has a superficial density that
  is about 6 times higher compared to our standard disk. As a consequence, when
  the planet decouples from the disk, the repeated crossing of the disk plane
  lead to a different dynamical evolution since the frictional force depends on
  the local disk density.
  The increase of only 25\% in the precession period of the disk initially
  extending over  $15\,\mbox{au}$, instead of the expected factor 2 on the basis of
  eq. 3), is due to the radial spreading of the disk due to the strong
  binary perturbations. The standard disk with initial outer
  radius of about  $30\,\mbox{au}$ cannot spread beyond the tidal truncation limit
  and, as a consequence, its outer radius remains constant. On the other hand, the
  smaller disk has room to expand in response to the binary gravitational
  perturbations extending after one cycle (about $2 \times 10^4$ yr) beyond $25\,\mbox{au}$.
  The outcome of the numerical simulations rule out warping as responsible for
  the decoupling of a planet from its disk in our dynamical configuration since, according
  to \citep{terquem2013}, a severe warping is needed to affect the planet inclination.
  However, this mechanism may be an important one for closer binaries, increasing
  the decoupling rate between planet and disk.

\begin{figure}[hpt]
    \resizebox{\hsize}{!}{\includegraphics{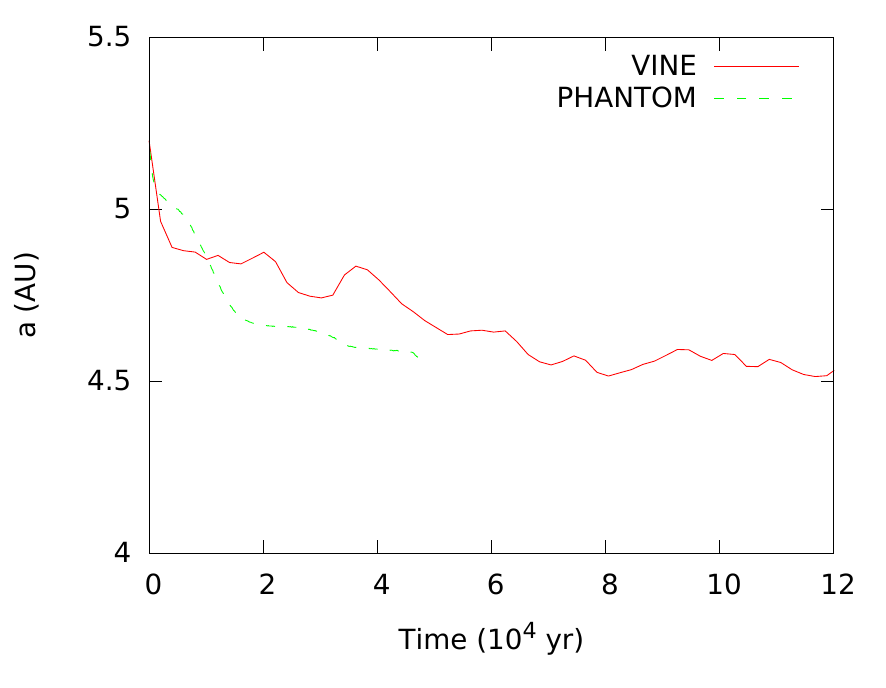}}
    \caption{\label{migr}
    Evolution of the semimajor axis of the planet in the \textsc{vine} and
    \textsc{phantom} runs, respectively.
    The planet migrates because of the friction with the gas when it periodically crosses
    the disk plane.}
  \end{figure}

  The planet migration, in Fig.~\ref{migr}, is significantly slower compared to that found
  by \citep[see][Fig.~3]{Xiang2014}.
  Apart from the initial fast migration rate when the planet is still embedded in the
  disk, when the planet detaches from the disk plane it is the friction developing during
  the periodic crossing of the disk by the planet that dominates the semimajor axis
  evolution and migration.
  According to \citet{teyssan2013}, the planet experiences a significant friction while
  crossing the disk due to aerodynamic drag and to a dynamical drag due to the
  gravitational scattering of the disk particles.
  The latter dominates since the ratio between the two friction forces can be expressed
  as \citep{teyssan2013}
  \begin{equation}
    \label{eq:friction}
    \frac{F_\mathrm{aer}}{F_\mathrm{dyn}} \sim \frac{1}{8\ln\left(\frac{H(r)}
    {R_\mathrm{p}}\right)}\left(\frac{R_\mathrm{p}}{a}\right)^2 \left(\frac{M_\mathrm{s}}
    {M_\mathrm{p}}\right)^2
  \end{equation}
  where $M_\mathrm{p}$ and $R_\mathrm{p}$ are the mass and radius of the planet,
  respectively, and $a$ is its semimajor axis.
  In our initial configuration where $a = 5\,\mbox{au}$, $H(r)= h_\mathrm{s} a \sim 0.19$
  and assuming a radius of the planet
  $R_\mathrm{p} \sim 2 R_\mathrm{J}\sim 0.001\,\mbox{au}$ we get a ratio lower than
  $0.02$.
  In this situation the drag force is mostly due to the particle scattering during the
  planet crossing.
  There may also be an angular momentum exchange between the planet and the disk when the
  planet approaches the disk, but it has probably less effect on the long term drift.

  The migration rate appears to be more irregular in the run with \textsc{vine} respect to
  that with \textsc{phantom}.
  This different behavior is possibly due to the mass accretion onto the planet in the
  \textsc{phantom} run (the planet is modelled as a sink particle) which it is not
  implemented in the \textsc{vine} run.
  In this last the planet cannot growth and gas is temporarily trapped and removed around
  it during the disk crossing possibly leading to a noisy migration.
  In the  run with \textsc{phantom}, the planet has reached a mass of $1.6$ $M_\mathrm{J}$
  after about $4 \times 10^4\,\mbox{yr}$  and then the mass growth is reduced.

\section{Initial mutual inclination of $60^{\circ}$}
\label{sec:init60}

\begin{figure}[hpt]
    \resizebox{\hsize}{!}{\includegraphics{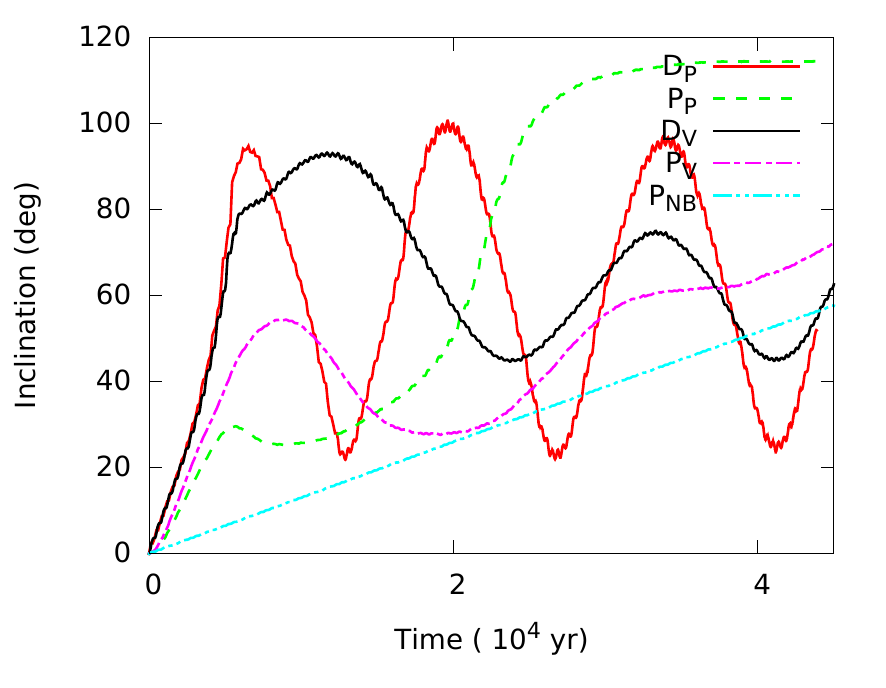}}
    \caption{\label{INC60}
    Evolution of the disk and planet inclination respect to the initial disk+planet plane
    in the case with $i_\mathrm{B} = 60^{\circ}$ in the \textsc{vine} and \textsc{phantom}
    runs, respectively.
    The symbols are the same as in Fig.~\ref{INC45f}.}
  \end{figure}

  Even in the scenario with $i_\mathrm{B} = 60^{\circ}$ we have run two different
  simulations, one with \textsc{vine} and one with \textsc{phantom}.
  In Fig.~\ref{INC60} we show the evolution of the disk and planet inclinations over a
  shorter timescale compared to the previous simulations where
  $i_\mathrm{B} = 45^{\circ}$.
  However, it is already clear that the planet abandons the plane of the disk and evolves
  independently under the action of the companion star in both simulations.
  The disk influence appears to be a perturbing force also in this different scenario.
  The planet inclination, as in the $i_\mathrm{B} = 45^{\circ}$, follows that of the
  disk only for a few $10^3\,\mbox{yr}$ after which the planet detaches from the disk
  following an N--body behavior while the disk perturbs the dynamics especially when the
  planet crosses the disk.
  In the two models the planet evolution is different and this can possibly be ascribed to
  the diverse way of handling the gas trapped around the planet and to the different
  behavior of the disks in the two cases.
  After an initial quick increase of the disk inclination, the growth halts at about
  $i_\mathrm{d} = 90^{\circ}$ and damped oscillations are thereinafter observed in both
  cases.
  However, the increase in inclination appears more irregular in the run with
  \textsc{vine} and stronger damped oscillations are observed immediately after the first
  maximum.
  In the run with \textsc{phantom}, the evolution appears more regular and the damping
  more progressive.
  This different behavior may be ascribed to the different viscosity which seems to play a
  more relevant role on the evolution of the disk inclination when $i_\mathrm{B}$ is
  higher.

  It is noteworthy that the damped oscillations recall, at some extent, those observed in
  \citet{martin2014}.
  However, the parameters of our model significantly differ from theirs so the exact same
  behavior is not expected.
  In particular, \citet{martin2014} use 1) different temperature and density profiles,
  2) their viscosity parameters are set in order to have a constant value of
  $\alpha_\mathrm{SS}$ all over the disk, and 3) their disk is ten times less massive
  compared to ours.
  As a consequence, the evolution of the disk inclination is only qualitatively similar,
  but it is interesting that in both scenarios damped oscillations are observed.
  An initial hint of damped oscillations of the disk inclination is also seen in the
  model with $i_\mathrm{B} = 45^{\circ}$ (Fig.~\ref{INC45f}) even if, as already observed,
  it appears feeble.

  A potential contribution to the strong damping in the $i_\mathrm{B} = 60^{\circ}$ case
  is possibly related to the exchange of disk mass between the two stars.
  After about $2 \times 10^4\,\mbox{yr}$ from the beginning of the simulation the
  secondary star has gained a gaseous disk whose mass is about $5\%$ that of the primary
  disk.
  The mass exchange is significantly lower in the $i_\mathrm{B} = 45^{\circ}$ case where
  the acquired disk has a mass of only $0.05\%$ of the initial disk mass.
  As a consequence, the viscosity of the disk plays a role also in an indirect way by
  forcing mass exchange for higher values of $i_\mathrm{B}$ leading to a damping of the
  disk inclination oscillations.

  \begin{figure}[hpt]
    \resizebox{\hsize}{!}{\includegraphics{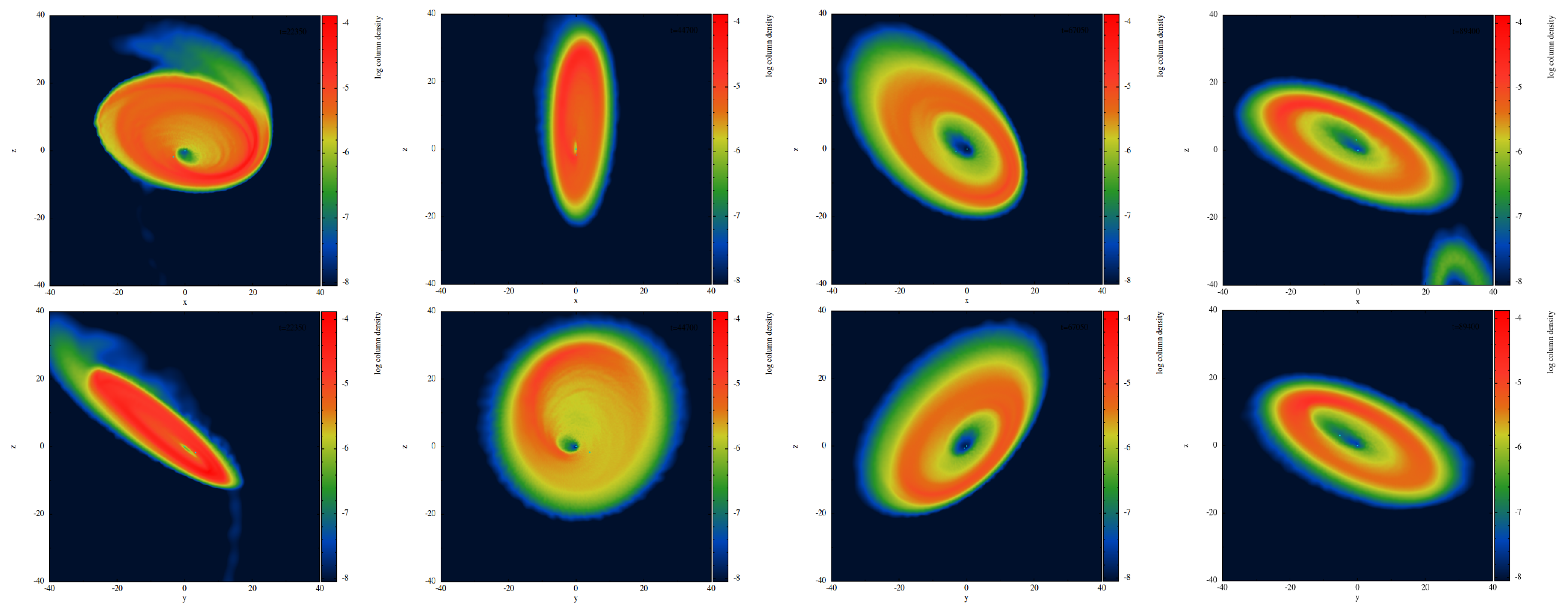}}
    \caption{\label{warp60}
    Evolution with time of the disk in the \textsc{phantom} model. Some mass leaves the
    primary disk and forms a ring around the secondary star. This behaviour was
    already observed in \cite{Picogna2013}. No significant warping is observed.
    }
  \end{figure}

   Even in this case we tested the potential warping of the disk by plotting the
   disk evolution every $300$ orbits of the planet. In this more inclined configuration
   we observe some mass transfer from the disk around the primary to the secondary star
   where a ring of gas slowly builds up.
   It is interesting to note that the
   disk eccentricity is higher compared to the case where $i_B = 45^o$ and comparable
   to that observed in \cite{martin2014}. This is why the mass flux from the
   primary towards the secondary star is more marked. In any case, the amount of
   mass extending out of the primary disk is not enough to cause a significant
   gravitational perturbation on the planet.

\section{Discussion and Conclusions}
\label{sec:concl}

  We have shown that a Jupiter--like planet embedded in a circumstellar disk will evolve
  almost independently from the disk in response to the perturbations of a misaligned
  binary stellar companion for values of the semimajor axis $a_B$ around
  $100\,\mbox{au}$.
  For inclinations of the binary plane of the order of $45^{\circ}$ and $60^{\circ}$,
  after an initial coupled evolution, the planet dynamically detaches from the disk and
  its inclination oscillates on a significantly longer timescale compared to that of the
  disk.
  This occurs since the N--body secular perturbations of the companion star dominate over
  the damping force of the disk which tends to drag the planet back into its median
  plane.
  Our simulations have $3$--times the resolution of those performed by \citet{Xiang2014}
  and were performed with $2$ different up to date SPH codes, \textsc{vine} and
  \textsc{phantom}.

  Our results show that one of the fundamental requirements of the model of
  \citet{Batygin2012} to explain the observed spin--orbit misalignment of some exoplanets
  is not met for small binary separations and a reasonable choice of physical parameters of the disk and
  planet.
  However, the migration of the planet appears to occur at a significant rate even if the
  orbit of the planet and the disk plane are not aligned.
  This is due to the friction with the gas during the repeated crossing of the disk by
  the planet.
  As a consequence, the inward drift of the planet would not be due to type I/II
  migration, as argued  for wider separations by \citet{cridabatygin2014},
  but to the dynamical friction with the
  disk.
  This would possibly lead to conclusions similar to those given in
  \citet{Batygin2012} concerning the long term evolution of the planet spin axis and
  semimajor axis migration, but the physical mechanism would be different.
  The formation of hot Jupiters with spin misaligned respect to the stellar equator
  would in this case not occur because the planet follows the disk in its precession, but by
  a combination of N--body Kozai cycles and repeated crossing of the disk plane by the
  inclined planet evolving out of the disk. Even if the disk oscillations are
  damped over a long timescale to a given value of inclination, the continuous crossing
  of the disk by the planet following the Kozai--Lidov cycles would grant
  a significant migration.

  Of course, a deeper exploration of the parameter space is needed to have a more complete
  scenario of the behavior of the system.
  We have also explored only one particular step of the system evolution while previous
  steps have also to be investigated.
  Ad example, when does the planet begin to evolve independently from its birth disk?
  At which stage of its growth? For which value of mass?

  We have shown in our simulations that a Jupiter mass planet exits almost immediately
  from the disk, but does this occur also for smaller mass planets?
  Possibly not, since the final infall of gas on the planet core is a fast process
  occurring on a timescale shorter than the decoupling between the planet and disk
  inclination.
  However, the problem of growing a planet core in a disk perturbed by an inclined binary
  is a critical process.
  The condensation of dust into larger pebbles and, eventually, kilometer--sized
  planetesimals may be strongly perturbed in particular when the planetesimals decouple
  from the gas.
  For small inclinations between the circumprimary gas disk and the binary orbital plane
  \citet{xie2009} and \citet{xie2010} showed that planetesimal accretion may be fast
  because size segregation among planetesimals favors low velocity impacts.
  For larger inclinations, the nodal longitude randomization forced by the companion star
  may lead to the dispersion of the planetesimal disk reducing the chances of planet
  formation.
  However, \citet{mar2009} showed that for binary semimajor axes larger than
  $70\,\mbox{au}$ the nodal longitude randomization timescale is longer than the
  planetesimal accretion process and core formation might occur.
  As a consequence, planet formation might occur within the disk even in presence of an
  inclined binary until the planet reaches a Jupiter mass or even more.
  After that, it would decouple from the disk and follow and independent evolution.

  Additional simulations are needed to better explore the problem and different initial
  values have to be sampled for the planet mass, the binary orbital elements, and the disk
  parameters (like mass, viscosity, temperature distribution, density profile and so on)
  which can influence the propagation of perturbations.
  Running more numerical models may help to understand not only the response on
  the long term of the disk to the binary perturbations and the possible occurrence of
  damping, but also the dynamical evolution of the planet.
  The problem with these type of simulations is the large amount of CPU required by any
  run of an SPH code in this configuration.

  Even if the migration of the planet in this scenario is due to the friction with the
  the gas of the disk, it remains an open problem how to drive the planet very close
  to the star.
  Isothermal simulations of disks in binaries \citep{kley2008,marza2009} show on the long
  term the formation of an elliptical hole in the density distribution close to the
  primary star.
  This would evidently halt the planet migration once the planet moves within the
  hole where either type I/II migration or, as we suggest in this paper, friction with
  the gas during the crossing of the disk would be switched off.
  However, more recent simulations with radiative disks do not show the formation of such
  an internal low density region \citep{kley2012,marza2012,Picogna2013}.
  This is an additional aspect that should be investigated to really understand the
  dynamical evolution of inclined hot Jupiters in misaligned binary star systems.

  \begin{acknowledgements}
    We thank an anonymous referee for his useful comments and suggestions.
     Many of our plots were made with the SPLASH software package \citep{Price2007}.
    G. Picogna acknowledges the support through the German Research Foundation (DFG)
    grant KL 650/21 within the collaborative research program ''The first 10 Million
    Years of the Solar System''.
    Some simulations were performed on the bwGRiD cluster in T\"ubingen, which is
    funded by the Ministry for Education and Research of Germany and the Ministry
    for Science, Research and Arts of the state Baden-W\"urttemberg, and the cluster
    of the Forschergruppe FOR 759 ''The Formation of Planets: The Critical First
    Growth Phase'' funded by the DFG.
  \end{acknowledgements}

\bibliographystyle{aa}
\bibliography{paper}

 \end{document}